\documentclass[aps,pre,twocolumn,superscriptaddress,noeprint]{revtex4-2}

\usepackage[latin1]{inputenc}
\usepackage{graphicx}
\usepackage{grffile}
\usepackage{amssymb}
\usepackage{amsmath}
\usepackage{bbold}
\usepackage{xspace}
\usepackage{dcolumn}
\usepackage{bm}
\usepackage{color}
\usepackage{float}
\usepackage[extra]{tipa}
\usepackage{MnSymbol}
\usepackage{braket}
\usepackage{dsfont} 

\setcitestyle{square,numbers}

\DeclareMathOperator{\Tr}{Tr}

\newcommand{\eg}[0]{e.g.\@\xspace}

\newfont{\tensy}{cmsy10}
\newcommand{\chem}[1]{{$\fontdimen16\tensy=3.0pt
    \fontdimen17\tensy=3.0pt \mathrm{#1}$}}

\newcommand{\ham}[1]{\hat{H}_{#1}}

\newcommand{\fan}[1]{\hat{c}^{\vphantom\dagger}_{#1}}
\newcommand{\fcr}[1]{\hat{c}^{\dagger}_{#1}}
\newcommand{\fden}[1]{\hat{n}_{#1}}

\newcommand{\fandiag}[1]{\hat{\gamma}^{\vphantom\dagger}_{#1}}
\newcommand{\fcrdiag}[1]{\hat{\gamma}^{\dagger}_{#1}}

\newcommand{\Q}[1]{\hat{Q}_{#1}}
\renewcommand{\P}[1]{\hat{P}_{#1}}

\newcommand{\qvec}{\vec{q}}

\newcommand{\kB}{k_{\text{B}}}

\newcommand{\im}{\mathrm{i}}

\newcommand{\Hc}{\mathrm{H.c.}}
\newcommand{\absolute}[1]{\left| #1 \right|}

\newcommand{\expv}[1]{\left\langle #1 \right\rangle}
\newcommand{\expvtext}[1]{\langle #1 \rangle}

\newcommand{\expvc}[1]{\left\llangle #1 \right\rrangle_{\qvec}}

\newcommand{\expvctext}[1]{\llangle #1 \rrangle_{\qvec}}

\newcommand{\U}[1]{{\hat{U}}(#1)}
\newcommand{\Udag}[1]{{\hat{U}}^\dagger(#1)}

\newcommand{\nnpairs}[1]{\langle #1 \rangle}

\begin{document}

\title{Real-time evolution of static electron-phonon models in time-dependent electric fields}

\author{Manuel Weber}
\affiliation{\mbox{Department of Physics, Georgetown University, Washington,
    DC 20057, USA}}
\affiliation{\mbox{Max-Planck-Institut f\"ur Physik komplexer Systeme, N\"othnitzer Str.~38, 01187 Dresden, Germany}}
\author{James K. Freericks}
\affiliation{\mbox{Department of Physics, Georgetown University, Washington,
    DC 20057, USA}}

\date{\today}

\begin{abstract}
We present an exact Monte Carlo method to simulate the nonequilibrium dynamics of electron-phonon models in the adiabatic limit of zero phonon frequency. The classical nature of the phonons allows us to sample
the equilibrium phonon distribution and efficiently evolve the
electronic subsystem in a time-dependent electromagnetic field for each phonon configuration.
We demonstrate that our approach is particularly useful for charge-density-wave systems
experiencing pulsed electric fields, as they appear in pump-probe experiments.
For the half-filled Holstein model in one and two dimensions, we calculate the out-of-equilibrium response of the current and the energy after a pulse is applied
as well as the photoemission spectrum before and after the pump.
Finite-size effects are under control for chains of $162$ sites (in one dimension) or $16\times 16$ square lattices (in two dimensions).
\end{abstract}


\maketitle


\section{Introduction}

Electron-phonon interaction plays an important role in strongly-correlated materials
in which it can give rise to
superconductivity, the formation of polarons,
or charge-density-wave (CDW) order.
Progress in ultrafast optical spectroscopy has opened up a path
to drive these materials out of equilibrium using strong light pulses
and probe the evolution of these phases or the emergence
of new phenomena directly in the real-time domain \cite{RevModPhys.83.471}.
A recent focus of pump-probe experiments has been on
CDW materials like \chem{TaS_2} \cite{PhysRevLett.97.067402, PhysRevLett.105.187401,Stojchevska177, Hane1400173, Vogelgesang:2018aa, Zongeaau5501, PhysRevLett.120.166401}
or the rare-earth tri-tellurides \cite{Schmitt1649, Zhou:2021aa}.
Pumping light into these materials can lead to a long-time ringing of
the CDW amplitude mode, as observed in angle-resolved
photoemission \cite{PhysRevLett.97.067402}, and even to long-lived
metastable phases showing a large change in conductivity \cite{Stojchevska177}.
While it is widely debated whether in these materials CDW order arises from the Peierls instability or from a purely electronic mechanism, phonons always play an important role in the out-of-equilibrium dynamics and  relaxation towards a steady state.

Theoretical modeling of these systems is often based on
time-dependent Ginzburg-Landau theory or Boltzmann equations,
which provide a phenomenological description of the CDW state.
Migdal-Eliashberg theory is very successful for phonon-mediated
superconductors but less reliable for CDW systems (because of the different ways these systems are screened).
To understand the microscopic details of these systems,
we need efficient numerical techniques
to solve the 
quantum many-particle problem out of equilibrium.
While for a realistic modeling
we have to consider
both electron-electron
and electron-phonon interactions,
the exact solution of the phonon part
turns out to be the biggest challenge for simulations.
For small clusters,
exact diagonalization has been performed
with classical \cite{PhysRevB.76.075105, doi:10.1143/JPSJ.79.034708} and quantum phonons \cite{PhysRevLett.109.176402}.
The density-matrix renormalization group (DMRG)
has been applied to one-dimensional (1D) systems
\cite{doi:10.1143/JPSJ.81.013701, PhysRevB.96.035154, 2019arXiv191101718S},
but so far simulations could only reach
short time scales on small lattices.
The major challenge for wave-function-based methods is the growing 
number of excited phonons
with time.
Already in thermal equilibrium, the unbound bosonic Hilbert space 
makes these methods less efficient than for purely electronic systems.
Quantum Monte Carlo methods can avoid this problem and recently made progress in determining the equilibrium phase diagrams of phonon-coupled systems in higher dimensions
\cite{PhysRevB.98.085405, Costa:2020aa,
PhysRevLett.122.077601, PhysRevLett.122.077602, PhysRevLett.126.107205,
PhysRevB.102.161108,
Li:2020aa, PhysRevLett.126.017601, 2021arXiv210205060C, 2021arXiv210208899G,
PhysRevB.103.L041105},
but the dynamical
sign problem prevents applications to the real-time domain.
In infinite dimensions, nonequilibrium dynamical mean-field theory \cite{PhysRevLett.97.266408, RevModPhys.86.779} gives exact results for the CDW phase in the purely electronic Falicov-Kimball model
\cite{PhysRevB.93.045110, PhysRevB.94.115167, PhysRevLett.122.247402},
whereas phonons can only be included approximately
using a strong-coupling impurity solver \cite{PhysRevB.88.165108, 2015EL....10937002W} or Migdal's approximation \cite{PhysRevB.91.045128}.
Approximate results can also be obtained from weak-coupling perturbation theory \cite{PhysRevX.3.041033, PhysRevB.87.235139, PhysRevB.90.075126}.
At this time, all available methods that solve the nonequilibrium electron-phonon problem exactly are restricted to small system sizes and short time scales,
inhibiting simulations for the experimentally-relevant case of quasi-2D systems.

In this article, we show that the electron-phonon problem driven by a
time-dependent electromagnetic field can be solved efficiently in the adiabatic
limit of infinite ion mass.
In this limit, the phonon frequency is zero and the lattice loses its dynamics.
As a result, the phonons become classical variables and their
thermal distribution function can be sampled using a
classical Monte Carlo method in combination with exact diagonalization of a
quadratic electronic Hamiltonian \cite{Michielsen1997,1996MPLB...10..467M}.
A similar approach had been applied to the double-exchange
model for the manganites where electrons couple to classical spins \cite{PhysRevLett.80.845}.
While previous work mainly concentrated on equilibrium properties, we
show that this class of models is well suited to study the response
of a minimal interacting system
to a pulsed electric field as it appears
in pump-probe scenarios.
Because for
each phonon configuration the full problem reduces to a noninteracting
system in a time-dependent field, the time evolution can be obtained
by iteratively diagonalizing the time-dependent Hamiltonian.
Using the example of the Holstein model, we demonstrate that our
approach can access time scales that are long enough
to reach a steady state after a pulsed electric field has been applied.
System sizes of $162$ sites for a 1D chain and $16\times16$ sites
for a 2D square lattice are sufficient to control finite-size effects.
Starting from different initial temperatures, we show results for  time-dependent observables like current and energy as a
pump field is applied, as well as the photoemission spectra before
and after the pump.

The adiabatic approximation of zero phonon frequency is often used for
modeling the equilibrium properties of CDW materials. In many of these materials,
the phonon mode that drives the Peierls instability went soft, so it only has
a few meV \cite{2016CRPhy..17..332P} in the ordered phase,
which is much smaller than the electronic energy scale,
and the adiabatic approximation turns out to be in good agreement
with experiments \cite{PhysRevMaterials.3.055001}.
Indeed, a classical treatment of the phonons is justified as long as the Peierls
gap and/or the temperature are larger than the phonon frequency \cite{Brazovskii1976}, which was recently confirmed by quantum
Monte Carlo simulations \cite{PhysRevB.98.235117}.
The application of strong electric fields in
pump-probe experiments raises the average energy of the system
to be much higher than the phonon frequencies, implying that
the adiabatic approximation should be accurate for pumped CDW systems
at short time scales. Our results in the adiabatic limit
capture the relaxation of the electronic subsystem from scattering
off the phonons after a pump was
applied. However, the adiabatic approximation will not
be able to describe the correct long-time behavior of real materials, as the phonons are
conserved quantities and cannot exchange energy with
the electronic subsystem. At the end of this article, we will
discuss how this limitation can be overcome by including
a classical dynamics for the phonons.

The paper is organized as follows. In Sec.~\ref{Sec:Model} we define
the Holstein model in a time-dependent electric field, in Sec.~\ref{Sec:Method}
we introduce the equilibrium and nonequilibrium formalism of
the Monte Carlo method, in Sec.~\ref{Sec:Results} we present
results for the 1D and 2D Holstein model, and in Sec.~\ref{Sec:Conclusions}
we conclude.

\section{Model%
\label{Sec:Model}}

The Holstein model \cite{HOLSTEIN1959325} is given by the Hamiltonian
\begin{align}
\label{eq:Ham_HS}
\hat{H}
  =
  &-J \sum_{\nnpairs{i,j}\sigma} \fcr{i\sigma} \fan{j\sigma}
  + \sum_i \left( \mbox{$\frac{K}{2}$} \Q{i}^2 + \mbox{$\frac{1}{2M}$} \P{i}^2 \right)
  \nonumber \\
  &+ g \sum_{i\sigma} \Q{i} \left( \fden{i\sigma} - \mbox{$\frac{1}{2}$} \right) \, .
\end{align}
The first term describes the nearest-neighbor hopping of electrons with amplitude $J$.
Here, $\fcr{i\sigma}$ ($\fan{i\sigma}$) creates (annihilates) an electron at 
site $i$ with spin $\sigma$. The second term represents local harmonic oscillators
with displacement operators $\Q{i}$ and momenta $\P{i}$. We define the
optical phonon frequency $\omega_0 = \sqrt{K/M}$ where $K$ is the stiffness constant and
$M$ the phonon mass. The third term couples the phonon  displacement $\Q{i}$
to the local electron density $\fden{i\sigma} = \fcr{i\sigma} \fan{i\sigma}$
via a constant $g$. We use the dimensionless coupling parameter
$\lambda = g^2/KW$ where $W$ is the bandwidth of the noninteracting electron system.
We have $W=4J$ for the 1D chain and $W=8J$ for the 2D square lattice.
We use $J=1$ as the unit of energy
and consider half filling with $\expvtext{\fden{i\sigma}}=1/2$.

In the following, we only consider the adiabatic limit  $M\to\infty$
which corresponds to $\omega_0 \to 0$ at fixed $K=M \omega_0^2$.
Then, the phonon momenta $\P{i}$ drop out of the Hamiltonian
and the displacement operators $\Q{i}$ can be replaced by classical variables $q_i$.
The ground state of the adiabatic Holstein model can be obtained from the variational principle.
At half filling, the ground-state energy $E(\qvec)$ for the 1D chain and for the 2D square lattice considered in this article is minimized
by the mean-field ansatz
$q_i = \Delta / g \, \cos(\mathbf{K} \cdot \mathbf{R}_i)$.
The periodic lattice distortion is accompanied by CDW order with ordering vector
$\mathbf{K}=\pi$ [$\mathbf{K}=(\pi,\pi)$]  for the 1D (2D) case
which corresponds to an alternating (checkerboard) pattern of the electronic density.
The single-particle gap $\Delta$ can be estimated self-consistently from the mean-field equations.
For any finite electron-phonon coupling $\lambda$ we get $\Delta > 0$
and the ground state is twofold degenerate
under $\Delta \to -\Delta$.
At zero temperature, the perfect dimerization
reduces the Holstein model to a single-particle Hamiltonian with a fully-occupied lower band and an empty upper band separated by the gap $\Delta$.
At finite temperatures, the adiabatic Holstein model remains an interacting problem that can be solved with the Monte Carlo method discussed in Sec.~\ref{Sec:Method}.
In 1D, long-range CDW order can only exist at zero temperature 
due to the spontaneous breaking of a discrete $\mathds{Z}_2$ lattice symmetry,
so that
any finite temperature leads to a disordered phase. However, the Peierls gap $\Delta$ is only fully filled in at a finite coherence temperature where short-range 
CDW correlations disappear \cite{PhysRevB.94.155150}.
On the square lattice, 
CDW order remains stable up to a critical temperature, above which one can
find a disordered phase \cite{PhysRevB.98.085405}. 
In both cases, CDW order is strongest at $\omega_0=0$.
While the CDW ground state of the 1D chain can be destroyed
at a critical $\omega_{0,\mathrm{c}}$, it is expected to remain stable
for the 2D square lattice  \cite{PhysRevB.98.085405}.
For further information on the effects of quantum lattice fluctuations
on the thermodynamic properties of the Holstein model, we refer
to Refs.~\cite{PhysRevB.98.235117, PhysRevB.98.085405}.

After preparing the Holstein model in a thermal state, we want to drive the system out of equilibrium
using a time-dependent classical electromagnetic field
\begin{gather}
\mathbf{E}(\mathbf{r},t)
	=
	- \bm{\nabla} \Phi(\mathbf{r},t)
	- \frac{1}{c} \frac{\partial \mathbf{A}(\mathbf{r},t)}{\partial t} \, ,
\\
\mathbf{B}(\mathbf{r},t)
	=
	\bm{\nabla} \times \mathbf{A}(\mathbf{r},t) \, .
\end{gather}
The electric field $\mathbf{E}(\mathbf{r},t)$ and the magnetic field $\mathbf{B}(\mathbf{r},t)$
can be represented in terms of the static potential $\Phi(\mathbf{r},t)$ and the vector
potential $\mathbf{A}(\mathbf{r},t)$. We use the temporal gauge
with $\Phi(\mathbf{r},t) = 0$. Then, the vector potential can be included in Eq.~(\ref{eq:Ham_HS})
via the Peierls substitution
\begin{align}
\label{eq:PeierlsSub}
\fcr{i\sigma}\fan{j\sigma} \to \exp \bigg[-\frac{\im e}{\hbar c} \int_{\mathbf{R}_i}^{\mathbf{R}_j} \mathbf{A}(\mathbf{r},t) \, d\mathbf{r} \bigg] \, \fcr{i\sigma}\fan{j\sigma} \, .
\end{align}
Here, $\mathbf{R}_i$ is the Bravais vector that points towards lattice site $i$.
In the following, we set $e=\hbar=c=1$.
We pump our system with a spatially-homogeneous but time-dependent electric field
\begin{align}
\mathbf{E}(t)
	=
	\mathbf{E}_0 \exp \left(- \frac{t^2}{2\sigma_\mathrm{p}^2} \right) \, \sin (\omega_\mathrm{p} t) \, .
\label{Eq:PumpField}
\end{align}
Here, $\mathbf{E}_0$ is the field amplitude, $\sigma_\mathrm{p}$ the pump width,
and $\omega_\mathrm{p}$ the pump frequency. We assume that the pump pulse
is centered at $t=0$ and we choose a field that has no dc component, consistent with optical pulse excitation. Our approach neglects some magnetic field effects generated near when the field is turned on and turned off, because this field does not satisfy Maxwell's equations. This approximation corresponds to describing optical light in a crystal as having vanishing momentum, which is a common approximation.

\section{Method%
\label{Sec:Method}}

Microscopic models of electrons
that only interact with classical degrees of freedom
can be solved efficiently using 
the Monte Carlo method of Ref.~\cite{PhysRevLett.68.1410}.
The method has been applied to a variety of scenarios, including the coupling
to adiabatic phonons in Holstein or Su-Schrieffer-Heeger models \cite{Michielsen1997,1996MPLB...10..467M},
localized electrons in the Falicov-Kimball model
\cite{PhysRevB.74.035109},
classical spins in double-exchange models \cite{PhysRevLett.80.845},
or $\mathds{Z}_2$ spins in effective models for Kitaev spin liquids \cite{PhysRevLett.113.197205}.

In the following, we first review the equilibrium formulation of the Monte Carlo method,
before we discuss the nonequilibrium formalism.

\subsection{Equilibrium formalism}

The Monte Carlo method described in this section applies to a generic Hamiltonian 
of the form
\begin{align}
\label{Eq:HamMethod}
\hat{H} = \ham{\mathrm{el}}(\qvec) + V(\qvec)
\end{align}
that can be split into a classical potential $V(\qvec)$
and a bilinear electronic part 
\begin{align}
\ham{\mathrm{el}}(\qvec)
=
\sum_{ij} \sum_{\sigma} \fcr{i\sigma} \, \mathcal{H}^{\phantom{\dagger}}_{ij}(\qvec) \,  \fan{j\sigma} 
\end{align}
coupled to classical degrees of freedom $\qvec$.
For simplicity in notation, we assume that the single-particle Hamiltonian 
$\hat{\mathcal{H}}(\qvec)$
is equal for all spin components.
The partition function of Hamiltonian (\ref{Eq:HamMethod}) takes the form
\begin{align}
\label{eq:Zeq}
Z
  =
  \int d\qvec \,  e^{-\beta V(\qvec)}
  Z_{\mathrm{el}}[\qvec]
  \, ,
\end{align}
where $Z_{\mathrm{el}}[\qvec] = \Tr \exp [-\beta(\ham{\mathrm{el}}(\qvec) - \mu \hat{N} )]$ is the grand-canonical
partition function of the electronic subsystem (for a specific configuration $\qvec$ of the classical variables), $\beta=1/\kB T$ the inverse temperature,
$\mu$ the chemical potential, and $\hat{N}$ the total particle-number operator.

Any expectation value of the full system,
\begin{align}
\expvtext{\hat{O}}
=
\int d\qvec \, W_\mathrm{eq}[\qvec] \, \expvctext{\hat{O}} \, ,
\end{align}
can be expressed as a weighted average over the equilibrium
distribution of classical variables $\qvec$,
\begin{align}
\label{weight_conf}
W_\mathrm{eq}[\qvec]
=
\frac{1}{Z}
e^{-\beta V(\qvec)}
Z_{\mathrm{el}}[\qvec] \, ,
\end{align}
and the noninteracting expectation value of the electronic subsystem
for a fixed configuration $\qvec$,
\begin{align}
\label{obs_conf}
\expvc{\hat{O}}
=
\frac{1}{Z_{\mathrm{el}}[\qvec]}
\Tr \left\{
e^{-\beta[\ham{\mathrm{el}}(\qvec)-\mu \hat{N}]} \, \hat{O}_{\qvec}
  \right\}
\, .
\end{align}
The classical variables $\qvec$ can be sampled from $W_\mathrm{eq}[\qvec]$ using the Metropolis-Hastings algorithm. We propose local updates $q_i \to q_i' = q_i + \Delta q$
which only change a single coordinate. For each Monte Carlo update,
we need to diagonalize
$\hat{\mathcal{H}}(\qvec)$ to obtain
\begin{align}
\label{Eq:diagHam}
\ham{\mathrm{el}}(\qvec)
	=
	\sum_{\alpha \sigma} \epsilon_\alpha \fcrdiag{\alpha \sigma} \fandiag{\alpha \sigma} 
	\, , \qquad
	\fan{i\sigma} = \sum_\alpha \Gamma_{i\alpha} \fandiag{\alpha\sigma} \, .
\end{align}
Here, $\Gamma_{i\alpha}$
is the transformation matrix from real-space coordinates to the energy eigenbasis with eigenvalues $\epsilon_\alpha$.
Unless necessary, we will suppress the $\qvec$ dependence
of the eigenbasis in the following.
With this, the Monte Carlo weight
\begin{align}
W_\mathrm{eq}[\qvec]
	=
	\frac{1}{Z}
	e^{-\beta V(\qvec)}
	\prod_{\alpha \sigma} \left\{ 1+ e^{-\beta (\epsilon_\alpha[\qvec]-\mu)} \right\}
\end{align}
is always positive and
the Metropolis acceptance probability
becomes
$
R(\qvec \to \qvec')
	=
	\min(1,W_\mathrm{eq}[\qvec']/W_\mathrm{eq}[\qvec])
$.
To obtain independent configurations,
we need to perform $\mathcal{O}(L)$ local updates.
Therefore, the algorithm scales as $\mathcal{O}(L^4)$.
To improve the sampling at low temperatures, we use an exchange Monte Carlo technique \cite{doi:10.1143/JPSJ.65.1604}.
For further details on our implementation, see Ref.~\cite{PhysRevB.94.155150}.

For each Monte Carlo configuration $\qvec$, observables are calculated
from Eq.~(\ref{obs_conf}). Because expectation values are taken with respect
to a quadratic Hamiltonian, Wick's theorem is valid for each $\qvec$.
Therefore, the computation of thermal averages only requires access to the equilibrium
density matrix
\begin{align}
\expvc{\fcr{i\sigma} \fan{j\sigma}} 
	= 
	\sum_\alpha \Gamma_{j\alpha} \,
	n_\text{F}(\epsilon_\alpha) \, \Gamma^\dagger_{\alpha i} \, .
\end{align}
Here, $n_\text{F}(\epsilon_\alpha) = 1/\{1+\exp [ -\beta(\epsilon_\alpha -\mu) ] \}$
is the Fermi-Dirac distribution.

\subsection{Nonequilibrium formalism}

The calculation of observables from Eq.~(\ref{obs_conf}) is not restricted to
the equilibrium case. Wick's theorem remains valid in the real-time domain
so that any 
 time-dependent correlation function can be obtained from
the greater and lesser Green's functions,
\begin{align}
\im \, G^{>}_{ij\sigma}(t,t';\qvec) &=  \expvc{\fan{i\sigma}(t) \, \fcr{j\sigma}(t')} \, , \\
-\im \, G^{<}_{ij\sigma}(t,t';\qvec) &=  \expvc{\fcr{j\sigma}(t') \, \fan{i\sigma}(t)} \, .
\end{align}
The time-dependent single-particle operators are defined in the Heisenberg picture,
\begin{align}
\fan{i\sigma}(t) =\Udag{t,t_0} \, \fan{i\sigma}(t_0) \, \U{t,t_0} \, ,
\end{align}
and the time-evolution operator from $t_0$ to $t$ is given by
\begin{align}
\U{t,t_0} = \hat{\mathcal{T}}_t \, \exp \bigg[-\im \int_{t_0}^t dt' \ham{}(t') \bigg] \, ,
\end{align}
where $\hat{\mathcal{T}}_t$ is the time-ordering operator.
To get access to $\U{t,t_0}$,
we discretize
the real-time evolution into small intervals $\Delta t$,
so that $\U{t,t_0}$ can be obtained sequentially via
\begin{align}
\label{Eq:TimeEvGroup}
\U{t,t_0} = \U{t,t-\Delta t}  \, \U{t -\Delta t,t_0} \, .
\end{align}
Then, the time-evolution operator for a single time step can be approximated as
\begin{align}
\label{Eq:UTrott}
\U{t,t-\Delta t}
	=
	\exp \big[ -\im \Delta t \, \ham{}(t) \big] + \mathcal{O}(\Delta t^2) \, .
\end{align}
Here, we choose $\ham{}(t)$ to be constant for the time step $\Delta t$.
We want to emphasize that the time evolution with Eq.~(\ref{Eq:UTrott})
is still exact if $\hat{H}(t)$ has a discretized time dependence.
Because the Hamiltonian is quadratic, 
the time evolution of the annihilation
operator,
\begin{align}
\fan{i\sigma}(t) =  \sum_j  \mathcal{U}_{ij}(t,t-\Delta t) \, \fan{j\sigma}(t-\Delta t) \, ,
\end{align}
is determined by the single-particle evolution operator
\begin{align}
\label{Eq:TimeEvOp_construct}
\mathcal{U}_{ij}(t,t-\Delta t) 
	=
	\sum_{\alpha} \Gamma_{i\alpha}(t) \, e^{-\im \Delta t \, \epsilon_\alpha(t)} \,\Gamma^\dagger_{\alpha j}(t) \, ,
\end{align}
which can be obtained from diagonalizing $\hat{\mathcal{H}}(\qvec,t)$ 
according to Eq.~(\ref{Eq:diagHam}).
Eventually, the greater and lesser Green's functions become
\begin{align}
\nonumber
\im \, G^{>}_{ij\sigma}(t,t';\qvec)
	&= 
	\sum_\alpha 
	\mathcal{U}_{i \alpha}(t,t_0)
	\left[ 1 - n_\text{F}(\epsilon_\alpha) \right]
	\mathcal{U}^\dagger_{\alpha j}(t',t_0)
	\, , \\
-\im \, G^{<}_{ij\sigma}(t,t';\qvec) 
	&=
	\sum_\alpha 
	\mathcal{U}_{i \alpha}(t,t_0) \,
	n_\text{F}(\epsilon_\alpha) \, 
	\mathcal{U}^\dagger_{\alpha j}(t',t_0)
	 \, ,
\label{Eq:GF_conf}
\end{align}
where $\mathcal{U}_{i \alpha}(t,t_0) = \sum_j \mathcal{U}_{i j}(t,t_0) \, \Gamma_{j \alpha}(t_0)$.
We assume that at the initial time $t_0$ the system is in thermal equilibrium.

The computational effort for calculating observables in the real-time domain strongly depends on the scenario for the time evolution. The simplest case corresponds
to a time-independent Hamiltonian where the evolution operator
$\mathcal{U}_{\alpha\alpha}(t,t_0)=\exp[-\im \left(t-t_0\right) \epsilon_\alpha ]$
is fully determined by the equilibrium eigenvalues. As a result, one can
obtain the equilibrium spectral functions for each Monte Carlo configuration
directly from the Lehmann representation.
This approach can also be combined with nonequilibrium Green's function techniques, \eg, to calculate the charge transport through the system when noninteracting leads are attached
 \cite{PhysRevB.99.155157}.
 Another simple scenario are parameter quenches where the Hamiltonian
 has to be diagonalized once before and once after the quench but eigenvalues
 are constant within each time domain \cite{PhysRevB.97.165107}.
 For a more generic time-evolution scenario, $\hat{\mathcal{H}}(t)$ has to be diagonalized
 for each step $\Delta t$ such that the computational effort for each measurement becomes
 $\mathcal{O}(L^3 N_t)$ where $N_t$ is the number of time steps.
 Evolving the system up to long time scales will quickly dominate the computational costs over the sampling of
 the equilibrium distribution.
 In this article, we demonstrate that the Monte Carlo method introduced above is  nonetheless very powerful for the time evolution with pulsed electric fields.
 Further simplifications can occur for periodic driving where Floquet theory applies \cite{2021arXiv210704096W}.

\subsection{Calculation of observables}

Any time-dependent single-particle observable can be calculated from the greater and lesser
Green's functions in Eq.~(\ref{Eq:GF_conf}).  We prepare our system at an initial time $t_0$ in the equilibrium eigenbasis labeled by the index $\alpha$. The time
dependence of Eq.~(\ref{Eq:GF_conf}) can then be accessed by propagating
$\mathcal{U}_{i\alpha}(t,t_0)$ forward in time. For this, we iteratively
diagonalize $\hat{\mathcal{H}}(t)$, set up the evolution operator for
a time step $\Delta t$ according to Eq.~(\ref{Eq:TimeEvOp_construct}),
and carry out the time evolution in Eq.~(\ref{Eq:TimeEvGroup}) via a matrix multiplication.

Equal-time observables can be calculated after each time step from $G^<_{ij\sigma}(t,t)$. In this article, we will consider the electronic energy
\begin{align}
E_\mathrm{el}(t) = 
 -\frac{J}{L} \sum_{i\sigma} \expv{
e^{-\im \phi_{i\ell}(t)}
\fcr{\mathbf{R}_i, \sigma}(t) \, \fan{\mathbf{R}_i + \mathbf{e}_\ell, \sigma}(t) + \Hc} 
\end{align}
as well as the total current $j(t) = \mathbf{j}(t) \cdot \mathbf{E}_0 / \absolute{\mathbf{E}_0}$ where
\begin{align}
j_{\ell}(t) = -\frac{J}{L} \sum_{i\sigma} \expv{\im \,
e^{-\im \phi_{i\ell}(t)}
\fcr{\mathbf{R}_i, \sigma}(t) \, \fan{\mathbf{R}_i + \mathbf{e}_\ell, \sigma}(t) + \Hc} 
\end{align}
is the component in $\ell \in \{x,y\}$ direction and $\mathbf{e}_\ell$ the corresponding translation vector.
Here, $\phi_{i\ell}(t)$ is the flux created by the site- and time-dependent vector potential in Eq.~(\ref{eq:PeierlsSub}).

To get access to the full Green's functions,
we tabulate $\mathcal{U}_{i\alpha}(t,t_0)$ for each time step and only
evaluate Eq.~(\ref{Eq:GF_conf}) when the time evolution is completed.
In this way, we can calculate the
photoemission spectrum \cite{PhysRevLett.102.136401}
\begin{align}
\nonumber
P(\omega,t_{\mathrm{probe}})
	\simeq - \im
	\int_{-\infty}^\infty &dt \int_{-\infty}^\infty dt' \, s(t) \, s(t')  \, e^{-\im \omega(t -t')} \\ 
	&\quad\times \frac{1}{L} \sum_{i\sigma} G^<_{ii\sigma}(t,t')
\label{Eq:PES}
\end{align}
from the local lesser Green's function.
We use a Gaussian envelope function for the probe pulse,
\begin{align}
s(t)
	=
	 \frac{1}{\sqrt{2\pi} \sigma_\mathrm{probe}} \,
	  \exp\left[-\frac{\left(t-t_\mathrm{probe}\right)^2}{2 \sigma_\mathrm{probe}^2}\right] \, ,
\end{align}
centered at time $t_\mathrm{probe}$ and with width $\sigma_\mathrm{probe}$.
It is sufficient to evaluate $G^<_{ii\sigma}(t,t')$ for $t,t'$ chosen from an interval
around $t_\mathrm{probe}$ and in the end perform the discretized Fourier transform.

\section{Results%
\label{Sec:Results}}

\subsection{Spinless Holstein model in 1D}

We first consider the 1D spinless Holstein model as it has become the
standard test case for numerical approaches to the electron-phonon problem.
We set $\lambda=0.5$ and restrict our discussion to the half-filled case, for which
the thermodynamic properties were studied in Ref.~\cite{PhysRevB.94.155150}.
For the real-time evolution, we apply the pulsed electric field defined in Eq.~(\ref{Eq:PumpField})
with $E_0 = 1.0$, $\sigma_\mathrm{p} = 5.0$, and $\omega_\mathrm{p}=1.0$.
We use time steps of $\Delta t = 0.1$ for which discretization effects
are smaller than the line widths in our plots.

\begin{figure}
  \includegraphics[width=\linewidth]{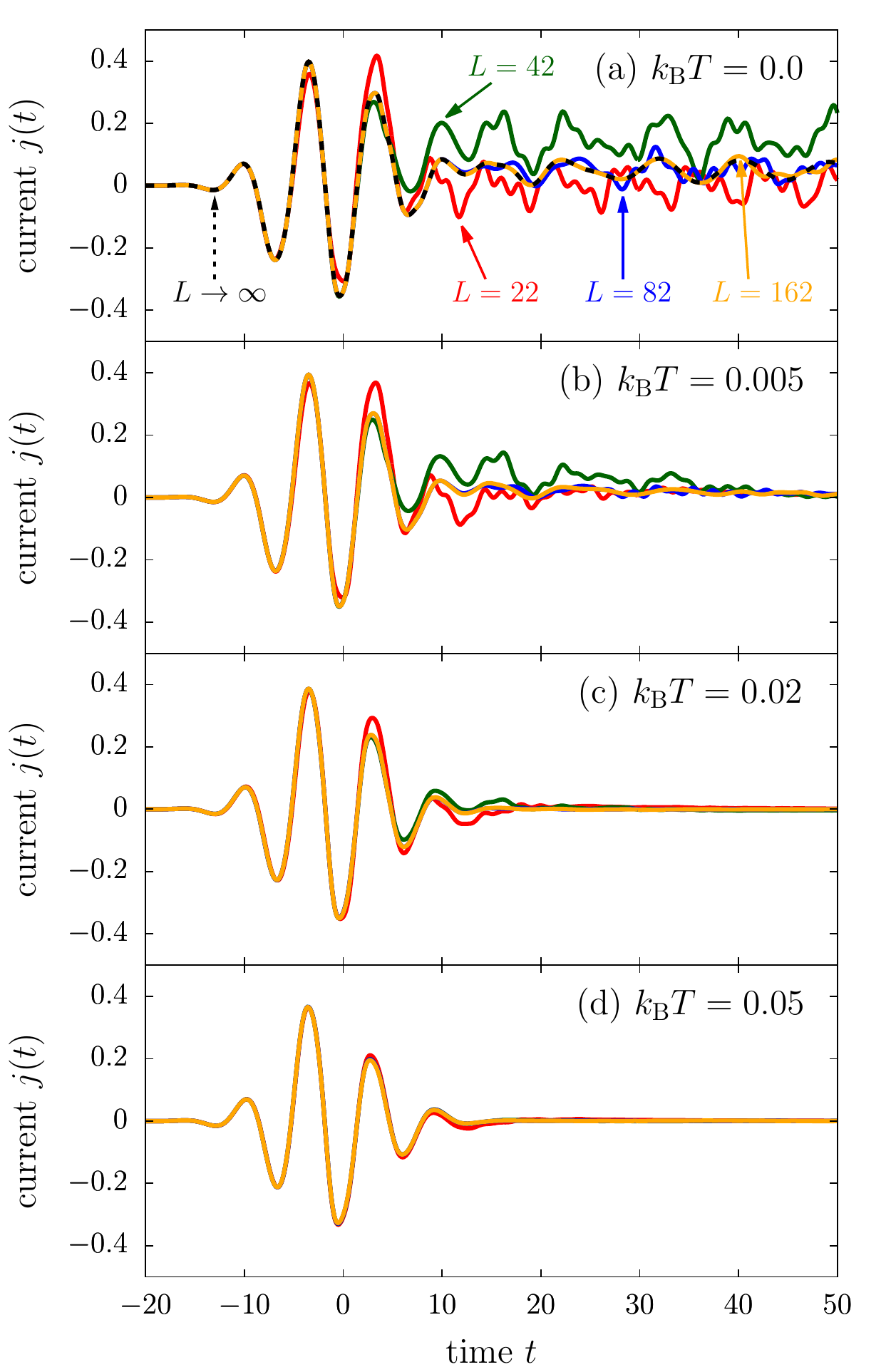}
  \caption{\label{fig:spinless_FS}
Finite-size analysis of the current in the 1D spinless Holstein model for different initial temperatures. Here, $\lambda=0.5$, $\Delta t = 0.1$. Error bars are smaller than the line width.
  }
\end{figure}
Figure~\ref{fig:spinless_FS} shows a finite-size analysis of the current for different initial temperatures. The effect of finite lattices is strongest at $\kB T=0$
in Fig.~\ref{fig:spinless_FS}(a) where the Holstein model
reduces to a noninteracting two-band problem via
the perfect
periodic lattice distortion $q_i = (-1)^i \Delta/g$.
For $\lambda=0.5$, we estimate the single-particle gap as $\Delta\approx0.34$.
At $\kB T =0$, the real-time evolution of the two-band system can be calculated efficiently \cite{PhysRevB.89.235129} such that we can get converged results in system size. Figure \ref{fig:spinless_FS}(a) shows that the onset of finite-size effects
can be delayed to longer times with increasing $L$. For $L=162$, the current
is converged up to $t=40$. Starting our simulations at finite temperatures substantially reduces finite-size effects, as demonstrated in Figs.~\ref{fig:spinless_FS}(b)--(d). The reduction of lattice-size effects
coincides with the suppression of short-range CDW correlations and the filling-in
of the Peierls gap at $\kB T \approx 0.1$. For 1D systems, we also find that finite-size effects
are stronger for pulsed electric fields than for the dc fields considered in Ref.~\cite{2021arXiv210704096W}.

\begin{figure}
  \includegraphics[width=\linewidth]{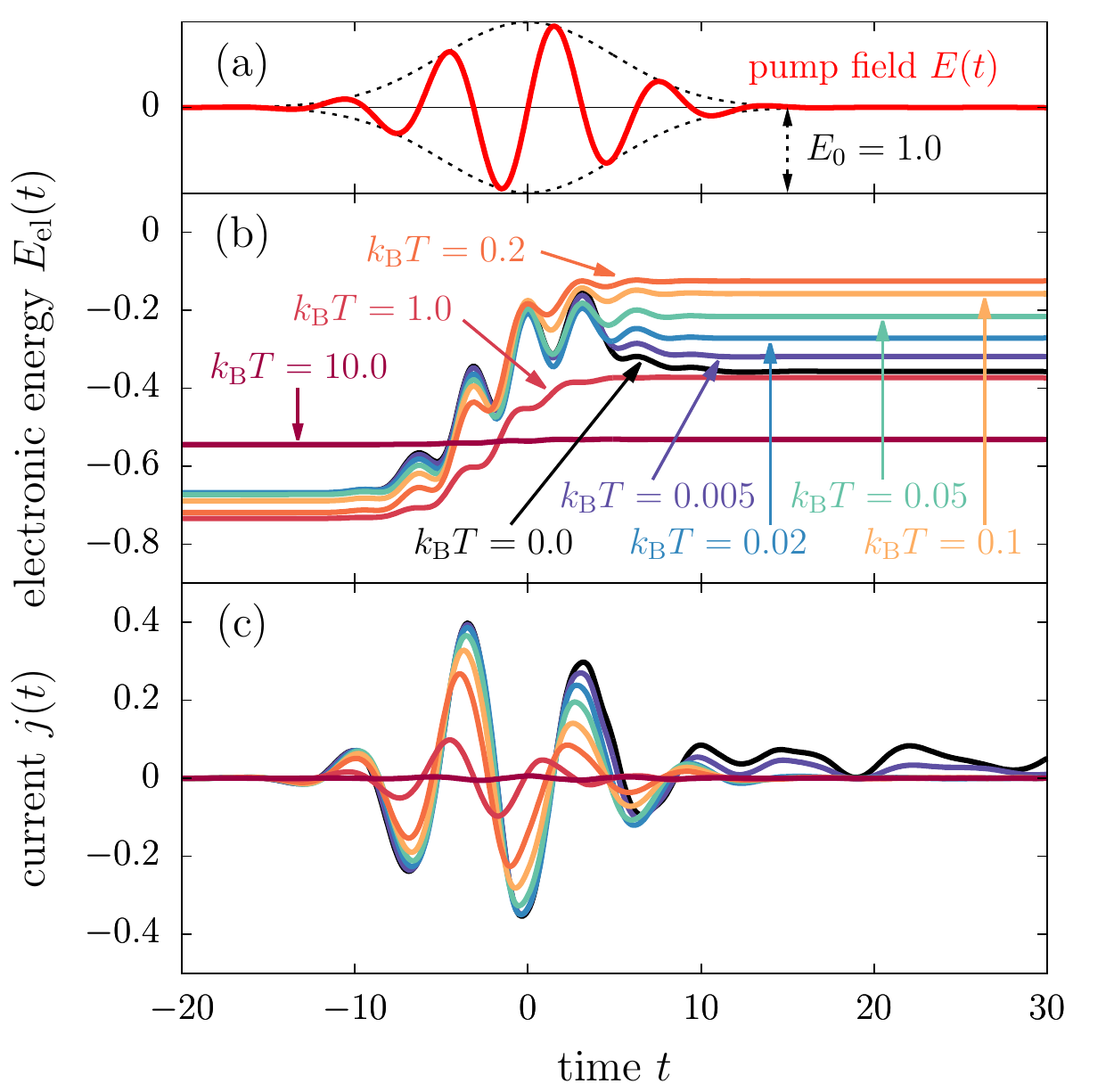}
  \caption{\label{fig:spinless_realtime}
Time-dependent response of the 1D spinless Holstein model for different initial temperatures.
For the pump field shown in panel (a), we plot (b) the electronic energy and (c) the current.
Here, $\lambda=0.5$, $L=162$, $\Delta t = 0.1$.
  }
\end{figure}
In Fig.~\ref{fig:spinless_realtime}, we study the effect of the initial temperature
on the electronic energy $E_\mathrm{el}(t)$ and the current $j(t)$. We set
$L=162$ for which finite-size effects are negligible. We find that
the oscillations of $E_\mathrm{el}(t)$ in Fig.~\ref{fig:spinless_realtime}(b) closely
follow the time dependence of the pump field in Fig.~\ref{fig:spinless_realtime}(a).
In the low-temperature regime, the initial energy only changes slightly
with increasing $\kB T$, whereas the final energy substantially increases with
$\kB T$. It therefore appears that the heating induced by the pump is more
effective in the presence of thermal fluctuations that have filled in the equilibrium Peierls gap. This is in contrast to the scenario of a constant applied field
where $E_\mathrm{el}$ gets closest to an infinite-temperature steady state for the lowest $\kB T$ \cite{2021arXiv210704096W}. Only at the highest temperatures, the final energy decreases again. This is in agreement with the phonon-induced disorder becoming stronger at higher $\kB T$.
Figure \ref{fig:spinless_realtime}(c) shows that the 
current is increasingly damped with higher $\kB T$. In particular, for any
$\kB T >0$ the current approaches zero in the long-time limit.

\begin{figure}
  \includegraphics[width=\linewidth]{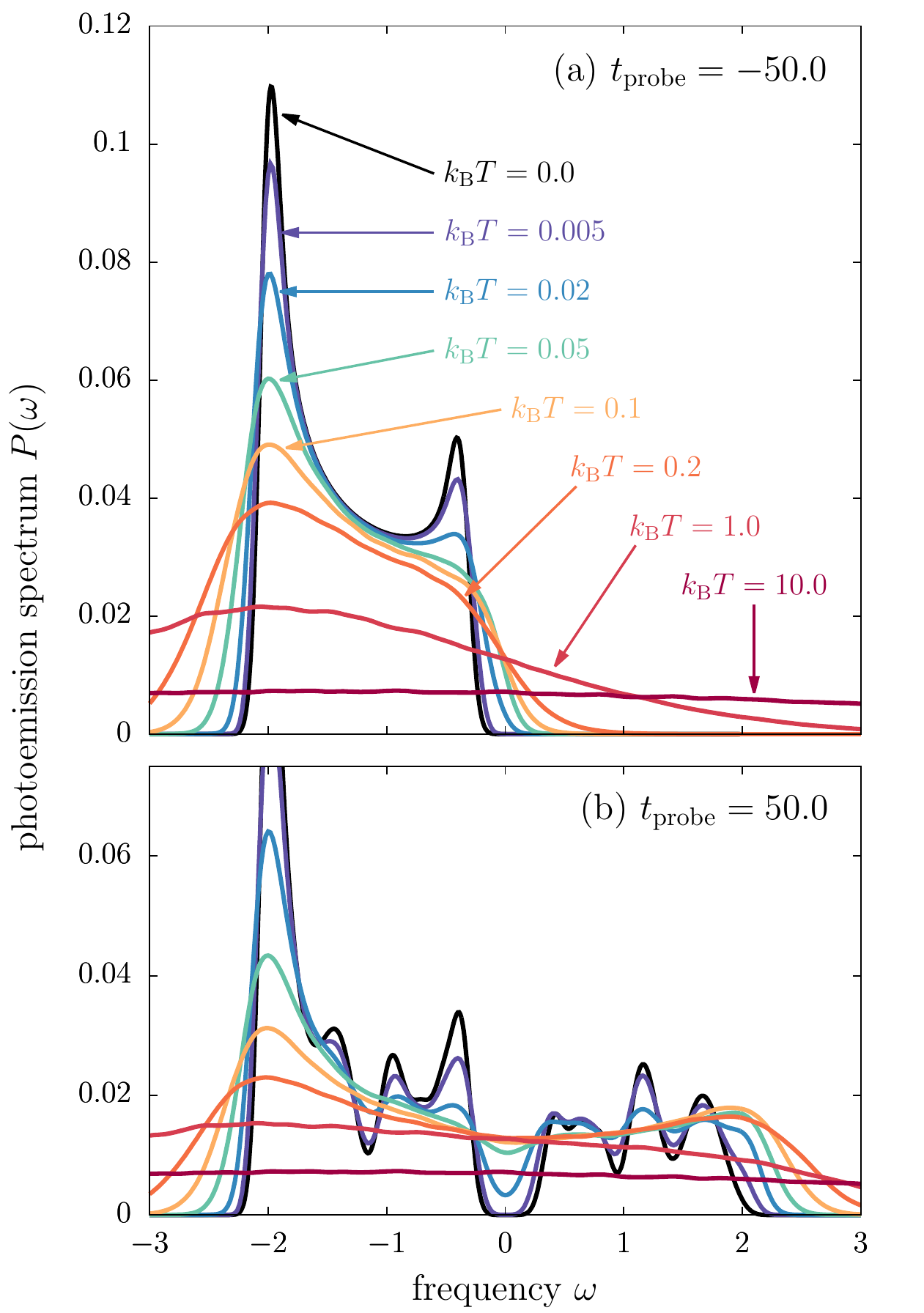}
  \caption{\label{fig:spinless_PES}
Photoemission spectrum of the 1D spinless Holstein model (a) before and (b) after the pump for different initial temperatures.
Here, $L=162$, $\lambda=0.5$, $\Delta t = 0.1$.
  }
\end{figure}
Figure~\ref{fig:spinless_PES} shows the photoemission spectra $P(\omega)$ before and after the pump field is applied,
as calculated from Eq.~(\ref{Eq:PES}).
We use a probe width of $\sigma_\mathrm{probe}=10.0$
to have high energy resolution and set $t_\mathrm{probe}=\pm 50.0$
for which the spectra have converged to their long-time limits $t\to \pm \infty$.
Figure \ref{fig:spinless_PES}(a) shows $P(\omega)$ in the initial state for different temperatures.
At $\kB T=0$, $P(\omega)$ probes the occupation of the lower band in the density of states. Note that the
square-root singularities at the band edges are smeared out by the finite probe width. Thermal fluctuations of the phonons
lead to a broadening of the spectral features and a closing of the single-particle gap \cite{PhysRevB.94.155150}. Because $P(\omega)$
probes the occupation of all available states according to the Fermi-Dirac distribution, we find an exponentially-activated tail
of excitations for $\omega>0$.
After the pump has been applied, the photoemission spectrum in Fig.~\ref{fig:spinless_PES}(b) includes substantial excitations in the upper band.
Our exact approach allows us to resolve the fine structure of the excitation spectrum, which includes several peaks in the upper band.
The main effect of the thermal fluctuations is the broadening of the sharp peaks that appear in the zero-temperature limit of perfect CDW order. Moreover, with increasing $\kB T$ the occupation of the highest accessible states at the upper band  edge increases.

\subsection{Spinful Holstein model in 2D}

In the following, we apply our method to the 2D spinful Holstein model on the square lattice.
At half-filling, the equilibrium problem has a thermal phase transition from a low-temperature CDW phase to a disordered phase 
that falls into the Ising universality class \cite{PhysRevB.98.085405}.
We consider $\lambda=0.15$ for which the mean-field gap at $\kB T=0$ is $\Delta \approx 0.55$ and
the critical temperature is $\kB T_\mathrm{c} \approx 0.1$ \cite{PhysRevB.98.085405}.
Starting from a thermal state, we drive our system with the pulsed electric field of Eq.~(\ref{Eq:PumpField})
applied in the diagonal direction $\mathbf{E}_0 = (1,1)$ with $\sigma_\mathrm{p} = 5.0$ and $\omega_\mathrm{p}=1.0$.
We use $\Delta t = 0.1$.

\begin{figure*}
  \includegraphics[width=\linewidth]{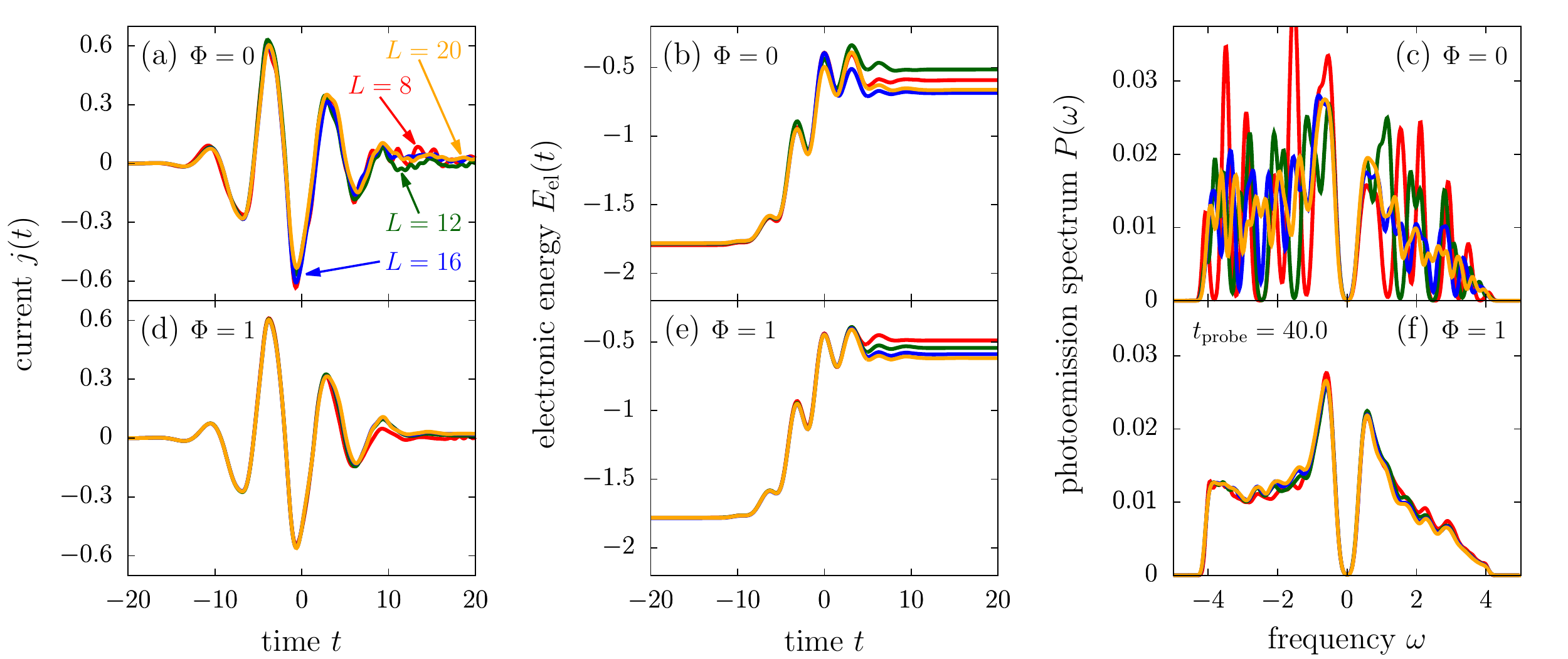}
  \caption{\label{fig:spinful_2D_FS}
Finite-size analysis of the current, electronic energy, and photoemission spectrum for the 2D spinful Holstein model on the square lattice. For each observable, we compare the size dependence of the plain system [(a)--(c)] with the situation where
one magnetic flux quantum $\Phi=1$ is threaded through the entire lattice [(d)--(f)]. Here, $\lambda=0.15$, $\kB T = 0.05$, $\Delta t = 0.1$.
  }
\end{figure*}
We start our discussion of the 2D Holstein model with a finite-size analysis of our nonequilibrium observables. Similar to the 1D case,
lattice-size effects are largest at $\kB T = 0$, where the half-filled Holstein model reduces to a noninteracting two-band problem. While thermal lattice fluctuations typically reduce lattice-size effects by broadening the delta excitations of a noninteracting system, tight-binding models on the square lattice usually suffer from strong size effects due to
 large level degeneracies. A common trick in equilibrium Monte Carlo simulations is to lift these degeneracies with
a static magnetic field that vanishes for $L\to \infty$ \cite{PhysRevB.65.115104}. Threading the entire lattice with one magnetic flux quantum
leads to an improved scaling already for small $L$. 
Figure~\ref{fig:spinful_2D_FS} compares the finite-size analysis of the current, electronic energy,
and photoemission spectrum in the CDW phase at $\kB T = 0.05$ with and without an applied magnetic flux $\Phi$.
Our simulations reach $L\times L$ lattices with $L\leq 20$. For $\Phi = 0$, the current
in Fig.~\ref{fig:spinful_2D_FS}(a) and the electronic energy in Fig.~\ref{fig:spinful_2D_FS}(b) show significant size
effects at the longest times and even at intermediate times $t\approx 0$ they are not yet converged with $L$.
By contrast, Figs.~\ref{fig:spinful_2D_FS}(d) and \ref{fig:spinful_2D_FS}(e) show quick convergence at $t\approx 0$
for $\Phi=1$. Moreover, $\Phi=1$ suppresses the finite-size oscillations of the current at the longest times and
leads to converged results for $L\simeq20$. The presence of the flux cannot fully eliminate the size effects of the final
electronic energy in Fig.~\ref{fig:spinful_2D_FS}(e), but it seems to enforce a monotonic dependence of $E_\mathrm{el}(L)$.
The photoemission spectra in Figs.~\ref{fig:spinful_2D_FS}(c) and \ref{fig:spinful_2D_FS}(f) show the largest improvement
when a flux is included. For $\Phi=0$, only the Peierls gap and the band width can be estimated reliably, whereas
the remaining spectrum is dominated by broadened delta peaks which only slowly evolve into well-defined bands with increasing $L$. For $\Phi=1$, a continuous spectrum that contains all the main features is already obtained for $L=8$. The fine structure of the lower and upper bands can be clearly identified with increasing lattice sizes.
All in all, the presence of a magnetic flux greatly reduces finite-size effects
in our nonequilibrium observables and enables a reliable estimation of the photoemission spectra.
Hence, we will use $\Phi=1$ for all results discussed below.

\begin{figure}
  \includegraphics[width=\linewidth]{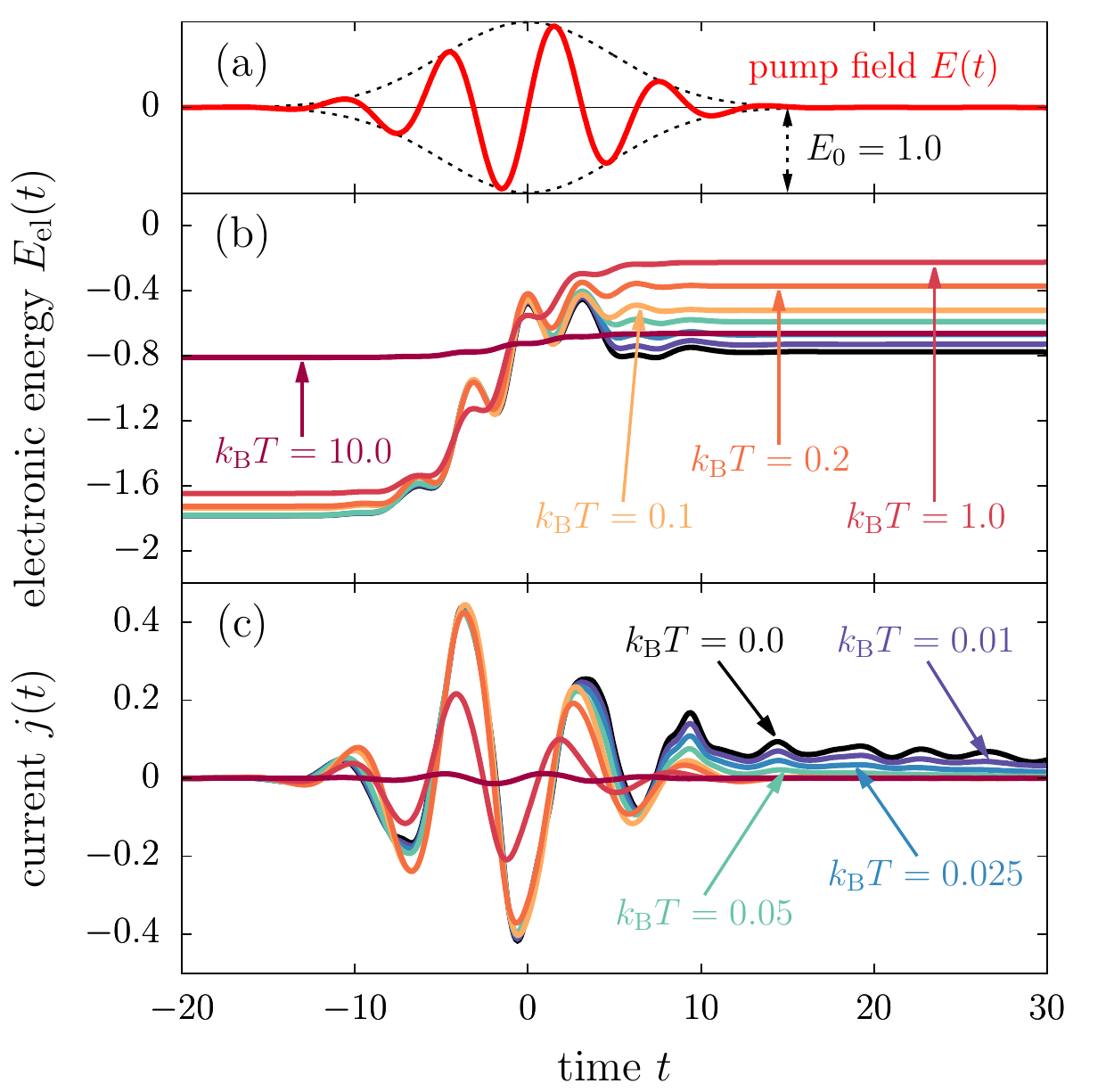}
  \caption{\label{fig:spinful_realtime}
Time-dependent response of the 2D spinful Holstein model on the square lattice for different initial temperatures.
For the pump field shown in panel (a), we plot (b) the electronic energy and (c) the current.
Here, $\lambda=0.15$,  $L=16$, and $\Phi = 1$.
  }
\end{figure}
Figure~\ref{fig:spinful_realtime} shows $E_\mathrm{el}(t)$ and $j(t)$ for different initial temperatures at $L=16$.
Our finite-size analysis suggests that the final energies in Fig.~\ref{fig:spinful_realtime}(b) might not be fully converged in $L$, but we expect
the relative changes with $\kB T$ to be consistent. Similar to the 1D case, the final electronic energy increases
with temperature up to $\kB T \approx 1.0$. Only for higher temperatures the thermal phonon disorder reduces the energy
absorption again.
As long as the pump field is applied, the current in Fig.~\ref{fig:spinful_realtime}(c) remains rather stable against thermal fluctuations if the initial state was in the CDW phase. Moreover, the long-time tail at $\kB T = 0$
is only weakly damped by thermal fluctuations. For the longest times considered in Fig.~\ref{fig:spinful_realtime}(c), the damping towards zero current only becomes stronger
as we approach $\kB T_\mathrm{c} \approx 0.1$.
If we initialize our system deep in the disordered phase, the current gets significantly damped also at intermediate times.
Interestingly, we observe a slight enhancement of $j(t)$ at early times for $\kB T = 0.1$ and $\kB T = 0.2$.
This might be related to the finite signal in the zero-frequency optical conductivity observed above the CDW transition in the Falicov-Kimball model, which has been interpreted as an effect of weak localization at small interaction strengths \cite{PhysRevLett.117.146601}.

\begin{figure}
  \includegraphics[width=\linewidth]{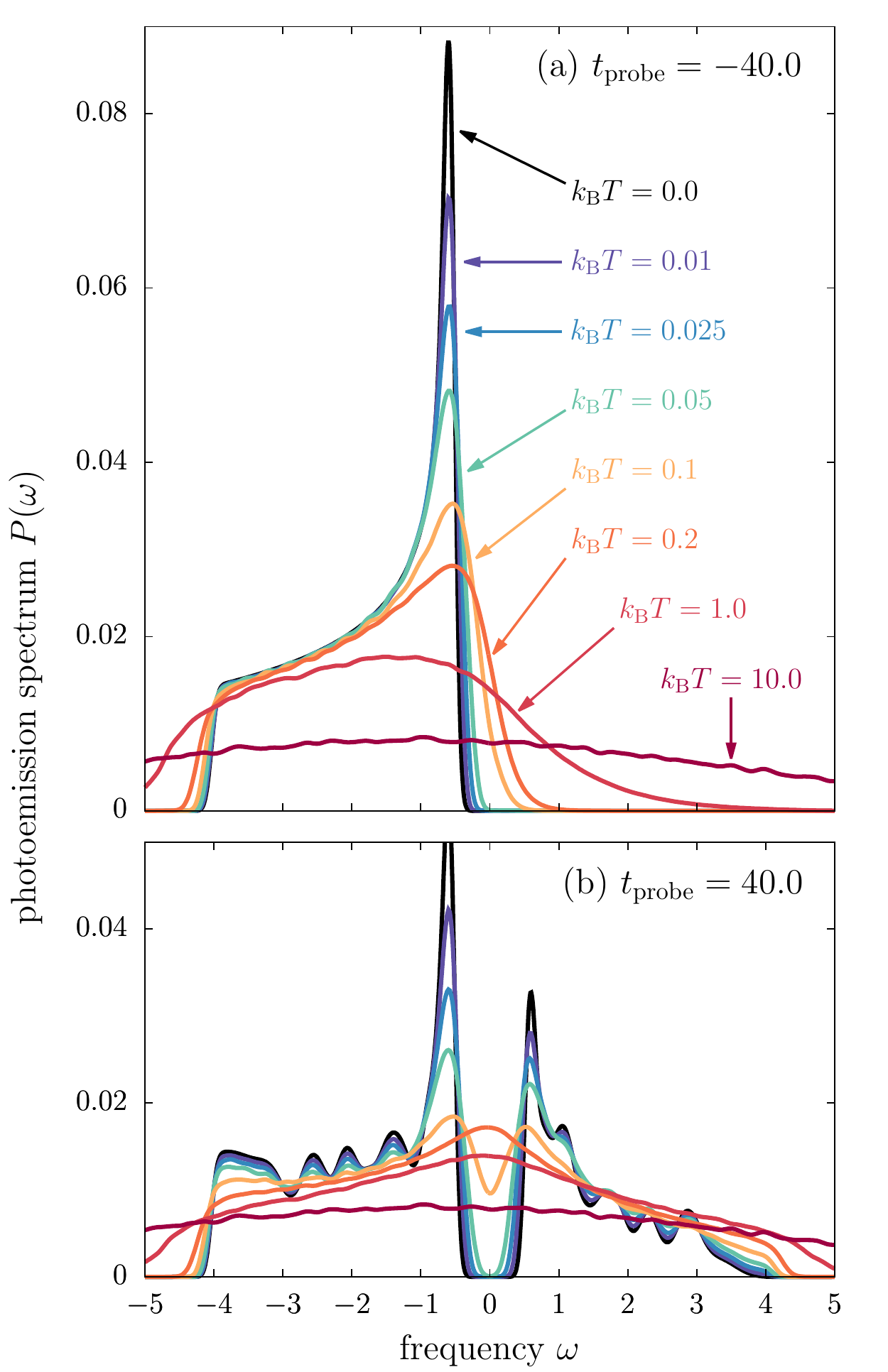}
  \caption{\label{fig:spinful_PES}
Photoemission spectrum of the 2D spinful Holstein model on the square lattice at $t_\mathrm{probe}=40.0$ (a) without and (b) with an applied pump field for different initial temperatures.
Here, $L=16$, $\lambda=0.15$, and $\Phi=1$.
  }
\end{figure}
Finally, Fig.~\ref{fig:spinful_PES} shows $P(\omega)$ for the 2D case before and after the pump is applied.
The noninteracting system has a van-Hove singularity at $\omega=0$ that is split by the Peierls distortion.
As already discussed for the 1D case, the occupation of the equilibrium spectrum in Fig.~\ref{fig:spinful_PES}(a) is governed by the Fermi-Dirac distribution. Application of the pump field leads to a broad range of excitations
in the upper band, as can be seen in Fig.~\ref{fig:spinful_PES}(b). In the CDW phase, we find well-defined
peaks in the upper and lower bands which get smeared out as the temperature reaches $\kB T_\mathrm{c} \approx 0.1$.
While the Peierls gap might disappear in the transient regime, it is recovered once the pulse is over.
Furthermore, we find that with increasing $\kB T$ spectral weight transfers towards the upper edge around $\omega=4.0$. Note that the low-amplitude oscillations in the spectra, especially at high $\kB T$, arise from the statistical fluctuations in the Monte Carlo data; the large amplitude oscillations in Fig.~\ref{fig:spinful_PES}(b) are a real effect.

\section{Conclusions \& Outlook%
\label{Sec:Conclusions}}

We have shown that the real-time evolution of electron-phonon models driven by a time-dependent
electromagnetic field can be calculated efficiently in the adiabatic limit. To this end, we used a classical
Monte Carlo method that samples the equilibrium phonon distribution and combined it with nonequilibrium
Green's function techniques. For each Monte Carlo configuration, we solved a noninteracting but time-dependent
electronic model with static phonon fields $\qvec$
as the phonons lose their dynamics in the adiabatic limit. This simplification allowed us to reach system sizes
of $162$ sites for a 1D chain and $16\times16$ sites for the 2D square lattice
 which is sufficient to control finite-size effects. 
 We demonstrated that size effects in nonequilibrium observables can be substantially reduced
 in the presence of a magnetic flux quantum threaded through the square lattice---a common trick
 in equilibrium Monte Carlo simulations \cite{PhysRevB.65.115104}. We presented results for the
 1D and 2D Holstein model driven by a time-dependent pump field. For different initial temperatures,
 we calculated the transient dynamics of the electronic energy and the current as well as the
 photoemission spectra before and after the pulse. We observed that thermal fluctuations
 enhance the system's ability to absorb energy from the pump. Moreover, the current
 is only slightly damped within the CDW phase of the 2D model, whereas phonon fluctuations
 lead to a stronger suppression of $j(t)$ in the disordered phase. Finally, we were able
 to resolve the fine structure in the photoemission spectra which appears in the CDW phase and
 gets smeared out by thermal fluctuations. All in all, we demonstrated that the classical Monte Carlo approach
 is well suited for studying driven quantum systems coupled to classical degrees of freedom. The formalism
 outlined in this paper not only applies to electron-phonon models but also to other types of interactions
 as they appear, \eg, in the Falicov-Kimball model or the double-exchange model.
 Further results on the driven electron-phonon problem will be presented elsewhere \footnote{R.~Nesselrodt, M.~Weber, J.~K.~Freericks, in preparation.}.
 
The adiabatic limit is expected to be a good approximation for short times after 
the pump has been applied. In this regime, the electrons can scatter off the thermally-induced phonon disorder
and relax towards a state with zero current. However, the static phonons prohibit the exchange of
energy between electrons and phonons which is crucial for the correct long time behavior observed in experiments.
To include these effects, one has to solve the full quantum phonon problem which is significantly
harder than the adiabatic limit discussed in this paper; indeed no algorithm is known that will work for large system sizes and long times on these types of problems. For example, exact diagonalization results
of the nonequilibrium problem are available on only up to $8$ lattice sites \cite{PhysRevLett.109.176402}.
A recent DMRG study of the 1D Holstein model driven far from equilibrium reached time scales
of $t \approx 6$ on $13$ sites \cite{2019arXiv191101718S}, whereas larger systems were obtained
for a weakly-driven system \cite{PhysRevB.96.035154}.
For both methods, the growing phonon occupation in the unbound bosonic
Hilbert space prohibits simulating to longer times. Moreover, DMRG works best at zero temperature and at high phonon frequencies
where the separation between the free-phonon energy levels is large \cite{2019arXiv191101718S}.
By contrast, the Monte Carlo method discussed in this paper works at $\omega_0 = 0$ and finite temperatures,
where larger system sizes and longer times can be obtained. It has the advantage that it is also applicable to 2D
where the out-of-equilibrium response can be studied across the finite-temperature Ising transition for these CDW-ordered systems.
We expect that our finite-temperature results remain accurate for low phonon frequencies, as long as $\kB T \gg \omega_0$.
To obtain a dynamical response of the phonons, it is possible to include an approximate molecular dynamics where the
phonons are propagated using their classical equations of motion \footnote{M.~D.~Petrovi\ifmmode \acute{c}\else \'{c}\fi{}, M.~Weber, J.~K.~Freericks, in preparation.}, as has been done for ground-state
simulations \cite{PhysRevB.76.075105, doi:10.1143/JPSJ.79.034708, PhysRevB.61.279}
and finite distributions for the initial phonon configurations \cite{PhysRevLett.96.086601, PhysRevX.10.021062}.

\begin{acknowledgments}
We thank F.~Assaad for helpful discussions.
 This work was supported by the U.S. Department of Energy (DOE),
  Office of Science, Basic Energy Sciences (BES) under Award
  DE-FG02-08ER46542. J.K.F.~was also supported by the McDevitt bequest at Georgetown University.
  The authors gratefully acknowledge the Gauss Centre for Supercomputing e.V. (www.gauss-centre.eu) for funding this project by providing computing time on the GCS Supercomputer SuperMUC-NG at Leibniz Supercomputing Centre (www.lrz.de) (project-id pr53ju).
\end{acknowledgments}

\end{document}